\documentstyle[12pt,preprint,aps,epsf]{revtex}

\firstfigfalse

\begin{document}
\preprint{hep-lat/9601001}
\title{
Chiral symmetry at finite $T$, the phase of the Polyakov loop
and the spectrum of the Dirac operator.}
\author{
M.A. Stephanov\thanks{E-mail address: \it misha@uiuc.edu.}}
\address{
	Department of Physics, 
	University of Illinois at Urbana-Champaign,\\
	1110 West Green Street, Urbana, IL 61801-3080, USA}
\date{January 1996}
\maketitle
\begin{abstract}
A recent Monte Carlo study of {\em quenched} QCD showed that
the chiral condensate is non-vanishing above $T_c$ in the phase where
the average of the Polyakov loop $P$ is complex.  We show how this is
related to the dependence of the spectrum of the Dirac operator on the
boundary conditions in Euclidean time. We use a random matrix model to
calculate the density of small eigenvalues and the chiral condensate
as a function of $\arg P$.  The chiral symmetry is restored in the
$\arg P=2\pi/3$ phase at a higher $T$ than in the $\arg P=0$ phase. In
the phase $\arg P = \pi$ of the $SU(2)$ gauge theory the chiral
condensate stays nonzero for all~$T$.
\end{abstract}

\newpage

\section{Introduction}

The relation between chiral symmetry breaking and confinement in gauge
theories has been a subject of continuous interest.  The picture of
chiral symmetry breaking suggested by Banks and Casher \cite{BaCa80}
relates the chiral condensate to the average density of small
eigenvalues of the Dirac operator in the fluctuating background of the
gauge field.  The appearance of small eigenvalues in turn can be
related to the fact that the QCD vacuum is populated by instanton
configurations \cite{Sh95}.

The pure QCD without quarks undergoes a phase transition at some
temperature $T_c$. It is interesting to see whether this transition
affects the behavior of small eigenvalues of the Dirac operator in
such a way that the chiral symmetry is restored. This motivates a
Monte Carlo study of QCD with quenched quarks. In this paper we
analyze the recent result \cite{ChCh95} of such a study. Unlike
previous similar works \cite{KoSt83} the authors looked at the value
of the chiral condensate in each of the $Z_3$ symmetric directions in
the complex plane in which the Polyakov loop acquires a nonzero
expectation value above $T_c$.  It was found that:
\begin{itemize}
\item[(a)] The chiral condensate $\langle \bar\psi\psi\rangle$
vanishes above $T_c$ in the phase with $\arg P=0$.
\item[(b)] In the phase with the $\arg P=2\pi/3$ the
$\langle \bar\psi\psi\rangle$ is nonzero for
some range of $T>T_c$\ !
\end{itemize}

The last point (b) contradicts a naive expectation that the vanishing
of the $\langle \bar\psi\psi\rangle$ has something to do with the
rearrangement of the QCD equilibrium state at $T_c$, e.g., a
suppression of the instanton configurations in the deconfined
phase. In this paper we show that this result can indeed be explained
very naturally if one takes into account {\em only} the effect of the
boundary conditions on the quark fields in the Euclidean
time. Nevertheless the fact (a) that the vanishing of $\langle
\bar\psi\psi\rangle$ coincides with $T_c$ tells us that the
rearrangement of the QCD equilibrium state at $T_c$ also plays a role.

\section{Polyakov loop and the spectrum of the Dirac operator}

The partition function of the QCD without quarks can be written as a
Euclidean path integral:
\begin{equation}\label{zs}
Z=\int {\cal D}A_\mu \exp\{-S[A]\};\qquad 
S=\int_0^{1/T} \int d^3\mbox{\boldmath $x$} F_{\mu\nu} F_{\mu\nu}.  
\end{equation}
An order parameter distinguishing the confined and the deconfined
phases is the expectation value of the Polyakov loop:
\begin{equation}
P(\mbox{\boldmath $x$}) = {1\over 3} \mbox{ Tr } \mbox{ P} \exp\left\{ig 
\int_0^{1/T}\, dt\, A_0(t,\mbox{\boldmath $x$})\right\},
\end{equation}
The value of $P$ is invariant under a gauge transformation periodic in
the Euclidean time: $U(1/T)=U(0)$. However, the action $S$ is also
invariant under aperiodic gauge transformations: 
\begin{equation}\label{uzu}
U(1/T)=zU(0),
\end{equation}
with $z=I\exp\{i2\pi/3\}$, which do not upset the boundary conditions
on $A_\mu$ because $z$ commutes with all matrices in the gauge
group. The value of the Polyakov loop acquires a phase
$\exp\{i2\pi/3\}$ after such a transformation. This corresponds to a
$Z_3$ symmetry transformation in an effective theory of the Polyakov
loops.

At small temperatures the expectation value of the Polyakov loop
is zero and the $Z_3$ symmetry is exact. Above $T_c$ the Polyakov
loop acquires a non-vanishing expectation value which can be related
to the liberation of color \cite{Po77}. 

\begin{figure}[hbt]
                \centerline{    \epsfysize=1.8in
                                \epsfbox{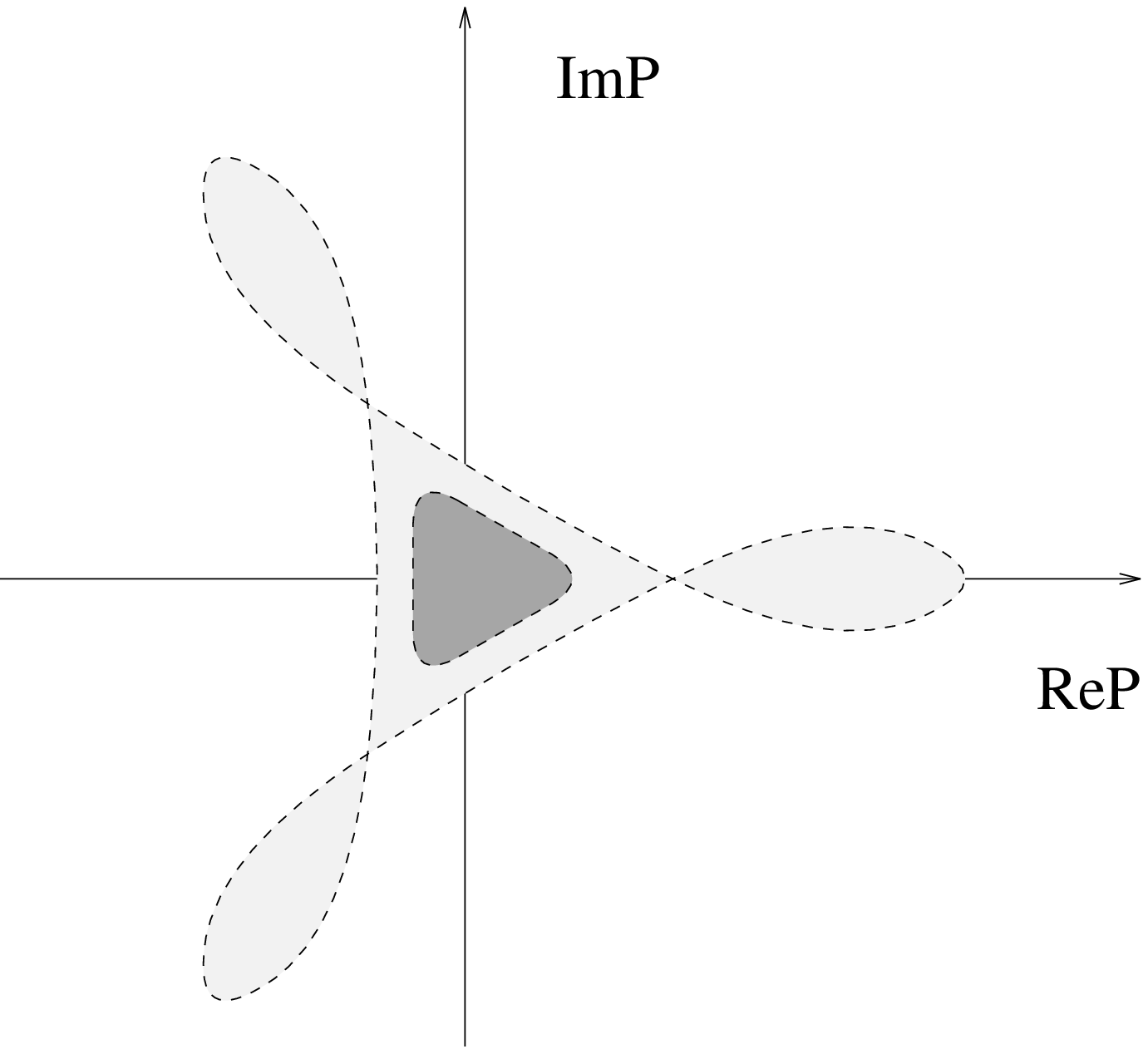}
                                \hspace{0.5in}
                                \epsfysize=1.8in
                                \epsfbox{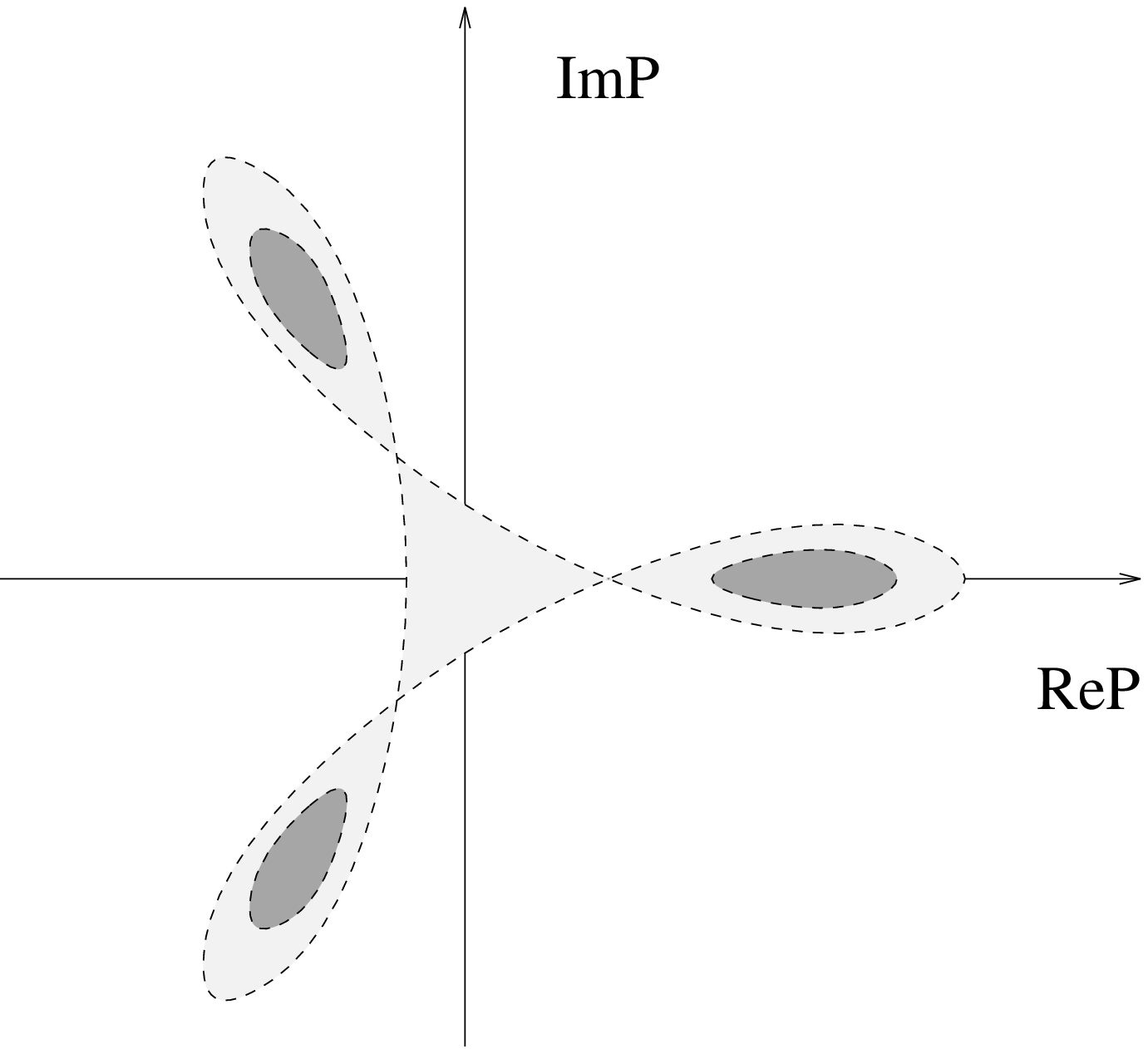}}
\hspace{160pt}(a)\hspace{180pt}(b)
\bigskip
\caption[]{
A schematic contour plot of the distribution of the
averaged value of the Polyakov loop: (a)~---~below $T_c$,
(b)~---~above $T_c$.  Only 2 isolines are drawn: one at the level of
the saddle point between the peaks and another in between the heights
of the peaks.}
\label{fig:z3}
\end{figure}

To analyze the results of the Monte Carlo simulation it is helpful to
consider a probability distribution of the Polyakov loop in a large
but finite volume. It can be calculated as a histogram \cite{hist}:
For each configuration one finds the averaged (over {\boldmath$x$})
$P$. The configuration then contributes a unit to the height of the
histogram at this $P$. The qualitative behavior near $T_c$ of this
distribution is shown in Fig.~\ref{fig:z3}. There are 4 peaks. At low $T$ the
peak at $P=0$ is taller. Most of the configurations have very small
$P$ near the origin. At $T=T_c$ the 3 degenerate peaks at nonzero $P$
are (roughly) of the same height as the peak at the origin. At higher
$T$ the $Z_3$ peaks are taller.

The chiral condensate is measured by averaging the inverse of the
Dirac operator over the gauge configurations. At high $T$ one has a
choice of averaging over ensembles of configurations clustered near
each of the $Z_3$ peaks. We refer to these ensembles as $Z_3$
phases. It was found in \cite{ChCh95} that the result is different in
the phases $\arg P=0$ and $\arg P=2\pi/3$.

Due to the $Z_3$ symmetry an ensemble of gauge configurations in the
phase $\arg P=2\pi/3$ can be mapped onto the ensemble at $\arg P = 0$
by an aperiodic gauge transformation (\ref{uzu}). The $Z_3$ phases of
pure QCD are equivalent. However, the spectrum of the Dirac operator
is not invariant under such a transformation.\footnote{A simple and
well known manifestation of this fact is the lifting of the degeneracy
between the $Z_3$ minima of the effective potential by the fermion
determinant contribution.}  We conclude that different behavior of the
chiral condensate in the two phases is entirely due to the effect of
the aperiodic gauge transformation on the spectrum of the Dirac
operator.

The action of the gauge transformation (\ref{uzu}) on the fermion
fields changes the boundary conditions in the Euclidean time
direction from antiperiodic: 
\begin{equation}\label{p-p}
\psi(1/T,\mbox{\boldmath $x$})=-\psi(0,\mbox{\boldmath $x$})
\end{equation}
to twisted: 
\begin{equation}\label{p-zp}
\psi(1/T,\mbox{\boldmath $x$})=-z\psi(0,\mbox{\boldmath $x$}).
\end{equation}
The eigenvalue problem for the Dirac operator is defined by the
differential equation: 
\begin{equation}\label{ev}
(D\hspace{-.6em}/ + m)\psi=\lambda\psi,
\end{equation}
with functions $\psi(x)$ satisfying given b.c. The b.c. in this problem
are different in the phases with $\arg P=0$ and $\arg P=2\pi/3$.

Moreover, qualitatively one can see that the effect of a nontrivial
$Z_3$ phase on the b.c.  tends to lower the eigenvalues, because of
less twisting required of $\psi$. It is transparent if you consider as
an example the free Dirac operator, i.e. $A_\mu=0$. In this case, the
spectrum in the phase $\arg P=0$ is given by:
$\lambda^2=\mbox{\boldmath $k$}^2 + ((2n+1)\pi T)^2$. While in the
$\arg P=2\pi/3$ phase it becomes: $\lambda^2=\mbox{\boldmath $k$}^2 +
((2n+1/3)\pi T)^2$. The smallest eigenvalue moves from
$\lambda=\pi T$ to $\lambda=\pi T/3$.

\section{Random matrix model}

To find the dependence of the spectrum on $\arg P$ for arbitrary
$A_\mu$ is a very difficult problem. Fortunately, the only information
we need about the spectrum is the density of small eigenvalues
$\rho(\lambda\to0)$ which is related to the chiral condensate by
\cite{BaCa80}:
\begin{equation}
\langle \bar\psi\psi \rangle = \pi \rho(0).
\end{equation}

To go further we need to make some approximations. As our first
approximation we neglect the Euclidean time dependence of the gauge
configurations responsible for small eigenvalues. This is not
altogether unreasonable approximation at least at $T$ of the order of
$T_c$ and higher for the following two reasons. First, it is the
infrared behavior of the gauge fields that is responsible for the
small eigenvalues. It is known, that the IR behavior of QCD at high
$T$ is given by a 3-dimensional theory \cite{GrPi81}. Second, the
correlation length at the first order phase transition in pure QCD may
be large compared to $1/T$ --- the extent in the time direction. For
the $SU(3)$ theory this is suggested by the fact that in the Potts
model, to which this transition is related by universality arguments,
the transition is very weakly first order. For the $SU(2)$ theory the
correlation length is infinite at $T_c$ and the theory {\em is}
effectively 3-dimensional at $T_c$.

\sloppy
In this approximation we can separate $t$ and {\boldmath$x$}
dependence in the eigenfunctions: $\psi(t,\mbox{\boldmath$x$}) =
\sum_n e^{i\omega_nt}\psi_n(\mbox{\boldmath$x$})$. Thus (\ref{ev})
reduces to the eigenvalue problem for a {\em three}-dimensional
operator:
\begin{equation}\label{ev3d}
[i\gamma_0\omega_n + i\gamma_0A_0
+ \mbox{\boldmath $\gamma D$} + m]\psi_n(\mbox{\boldmath$x$})
= \lambda\psi_n(\mbox{\boldmath$x$})
\end{equation}
for each $\omega_n$, where the gauge fields $A_0$, {\boldmath$A$}
depend only on {\boldmath$x$}. The values of $\omega_n$ are determined
by the b.c. in the Euclidean time:
\begin{equation}
\omega_n=((2n+1)\pi - \arg P) T. 
\end{equation}

\fussy
To find the density of small eigenvalues (or, chiral condensate) for
the operator (\ref{ev3d}) we use another approximation known as the
(Gaussian) random matrix model \cite{ShVe93}. We choose a basis
$\psi_a$ of $N$ functions\footnote{In 4d the functions $\psi_a$ can
(but need not necessarily) be thought of as the exact zero modes of
each of $N$ individual instantons and $X$ being the overlap matrix. In
3d a similar role can, perhaps, be played by periodic instantons or
dyons.}  (the same for all gauge field configurations and all
$\omega_n$) and write the {\em three\/}-dimensional Dirac operator in
this basis:
\begin{equation} 
\left (
\begin{array}{cc} 
m & iX+i\omega_n \\ 
iX^\dagger +i\omega_n & m
\end{array} \right ),
\end{equation}
where the $X$ is a complex $N\times N$ matrix. Then one can replace
the averaging over the gauge fields by averaging over the elements of
the matrix $X$. The main approximation of the model, which proves to
be very good (and, probably, exact in some sense) in 4d \cite{ShVe93},
is that the probability distribution is Gaussian. In other words, we
calculate:
\begin{equation}\label{z}
Z = \int {\cal D} X 
\exp\left\{-{NC^2} \,{\rm Tr}\, X X^\dagger\right\}
\prod_n {\det}^{N_f} \left (
\begin{array}{cc} 
m & iX+i\omega_n \\ 
iX^\dagger +i\omega_n & m
\end{array} 
\right ).
\end{equation}
This is actually an approximation to the full partition function
of QCD with $N_f$ flavors of {\em dynamical\/} fermions. We will be
interested, however, in the limit $N_f\to0$ --- quenched fermions.
In fact, one can check that $N_f$ cancels out in our final results
anyway. Therefore, to simplify the formulae we shall put $N_f=1$.
It is also convenient to view each Matsubara mode as a flavor with a
Matsubara mass $\omega_n$.
From $Z$ we can obtain the chiral condensate in the thermodynamic
limit:
\begin{equation}\label{pbpz}
\langle\bar\psi\psi\rangle = \lim_{m\to0} \lim_{V_3\to\infty} 
{T\over N_f V_3} {\partial\over \partial m}\ln Z.
\end{equation}
where $V_3$ is the volume of the 3-space.

The calculation \cite{ShVe93,JaVe95} of (\ref{z}) is similar in spirit
to the mean field calculation in the large~$N$ sigma models
(Hubbard-Stratonovich transformation).  We introduce a new random
matrix~$Y$, the integral over which can be calculated in the saddle
point approximation at large $N$. The first step is to rewrite the
determinant as a Grassmann integral:
\begin{equation}
Z =
\int {\cal D} X {\cal D} \bar\Psi {\cal D} \Psi
\exp\left\{-NC^2 \mbox{ Tr } XX^\dagger\right\}
\exp\left\{\sum_n \bar\Psi_n \left (
\begin{array}{cc} 
m & iX + i\omega_n \\ 
iX^\dagger + i\omega_n & m
\end{array}           \right )
\Psi_n  \right\}.
\end{equation}
The spinors $\Psi^a_n$ carry an implicit index $a$ ($a=1,\ldots,N$). 
Integration over $X_{ab}$ can be carried out 
which creates the four-fermion term:
\begin{equation}
-{1\over NC^2} \bar\Psi^a_{Ln}\Psi^b_{Rn}\bar\Psi^b_{Rm}\Psi^a_{Lm}.
\end{equation}
This term can be rewritten using another auxiliary random matrix $Y_{mn}$:
\begin{equation}
Z=\int {\cal D}Y {\cal D}\bar\Psi {\cal D}\Psi 
\exp\left\{-NC^2 \mbox{ Tr } YY^\dagger\right\}
\exp\left\{\sum_{a=1}^{N}\bar\Psi^a \left (
\begin{array}{cc}
Y+m & iM \\
iM & Y^\dagger + m
\end{array}           \right )
\Psi^a               \right\},
\end{equation}
where $M$ is a diagonal matrix $\omega_n\delta_{nk}$. 

The integral over Grassmann variables can be then performed:
\begin{equation}\label{zy}
Z=\int {\cal D}Y 
\exp\left\{-NC^2 \mbox{ Tr } YY^\dagger\right\}
{\det}^N \left( 
\begin{array}{cc}
Y + m & iM \\
iM & Y^\dagger + m
\end{array} \right)
.
\end{equation}

At large $N$ the saddle point equation for the matrix $Y$ is:
\begin{equation}\label{saddle}
- C^2 Y + (Y+m)
\left[ (Y^\dagger+m)(Y+m) + M(Y+m)^{-1}M(Y+m)\right]^{-1} = 0.
\end{equation}
This has a solution in the form of a diagonal matrix with eigenvalues
$y_n$ that satisfy (we put m=0 here):
\begin{equation}\label{gap}
C^2 y_n = {y_n\over \omega_n^2 + y_n^2}.
\end{equation}
This equation is analogous to the well known gap equation in the
large $N$ sigma models. A nontrivial solution $y_n=(1/C)
\sqrt{1-\omega_n^2C^2}$ exists only for those $n$ for which:
$1-\omega_n^2C^2>0$.

Using (\ref{pbpz}) and (\ref{zy}) we find that the chiral 
condensate is given by:
\begin{equation}\label{pbptr}
\langle\bar\psi\psi\rangle=\lim_{m\to0}\lim_{V_3\to\infty}
{T\over V_3} \left\langle N \mbox{ Tr } 
\left(
\begin{array}{cc}  
Y+m & iM \\
iM & Y^\dagger + m
\end{array} \right)^{-1} \right\rangle.
\end{equation}
In the saddle point approximation at large $N$ using (\ref{gap})
we find:
\begin{equation}\label{pbp}
\langle\bar\psi\psi\rangle = n_3
C \sum_n \sqrt{1-\omega_n^2C^2}\cdot\theta(1-\omega_n^2C^2),
\end{equation}
where $n_{3}=2N/V_3$ and 
$\omega_n=((2n+1)\pi-\arg P)T$.

We see that the chiral condensate vanishes (continuously: $\beta=1/2$
in this approximation) when $\min_n|\omega_n|>1/C$, in other words,
when the temperature exceeds:
\begin{equation}\label{tchi}
T_\chi(\arg P)= {1\over C|\pi - \arg P|},
\end{equation}
where $\arg P$ is mod $2\pi$.\footnote{In principle, $C$ depends on
$T$. For example, for $T\gg T_c$ one should expect $C^{-1}\sim g^2T$.
The equation~(\ref{tchi}) then defines $T_\chi$ self-consistently.  It
is not unreasonable to expect, though, that $C\approx {\rm const}$
near $T_c$.}  This result is in accordance with our expectation that
nonzero $\arg P$ favors chiral symmetry breaking.

Our model at $\arg P=0$ is similar to a model suggested in
\cite{JaVe95} where only the lowest Matsubara frequency was taken into
account. This makes sense since we are interested in the behavior of
small eigenvalues of the Dirac operator. Indeed, only the band
$\omega_0$ of the eigenvalues contributes in (\ref{pbp}) when $T$ is
close to $T_\chi$ and $0<\arg P<2\pi$. In our case keeping all
Matsubara frequencies ensures periodicity in $\arg P \bmod 2\pi$.

\section{Discussion and conclusions}

We analyzed a recent Monte Carlo observation that the chiral
condensate $\langle\bar\psi\psi\rangle$ does not vanish in some
interval above $T_c$ in quenched QCD in the phase with $\arg
P=2\pi/3$. This fact suggests that the rearrangement of the QCD
equilibrium state at $T_c$ (deconfinement) is not the only source of
the restoration of the chiral symmetry in quenched QCD. 

We calculated (approximately) the chiral condensate at finite $T$
taking into account the dependence of the spectrum of the Dirac
operator on the b.c. for the quark fields in the Euclidean time.  We
found that the $\langle\bar\psi\psi\rangle$ vanishes continuously at a
temperature $T_\chi(\arg P)$ which depends on the phase of the
Polyakov loop (\ref{tchi}).

This suggests the following interpretation of the result
\cite{ChCh95}. As we approach the deconfinement transition from
$T>T_c$ side the chiral symmetry breaking in the $\arg P=2\pi/3$ phase
occurs earlier, at a temperature $T_\chi(2\pi/3) > T_c$. The chiral
condensate is zero in the $\arg P=0$ phase because
$T_\chi(0)<T_\chi(2\pi/3)$.  It turns out that
$T_c>T_\chi(0)$. Indeed, the result~\cite{ChCh95} suggests that $T_c$
is only about 7\% lower than $T_\chi(2\pi/3)$, while in our
approximation $T_\chi(0)$ is about $1/3$ of $T_\chi(2\pi/3)$. At $T_c$
the $Z_3$ phases become metastable and at lower temperatures the
single $P=0$ phase dominates (see Fig.~\ref{fig:z3}).

Being cavalier enough, one can conjecture that the
effect of this transition on $\langle\bar\psi\psi\rangle$ can be taken
into account very roughly by averaging over 3 values of $\arg P$. This
would lead to a relation between the condensate just below $T_c$ and
the condensate in the $\arg P=2\pi/3$ phase just above $T_c$:
\begin{equation}
\langle\bar\psi\psi\rangle(T_c-)\approx{2\over3}
\langle\bar\psi\psi\rangle(T_c+, \arg P=2\pi/3),
\end{equation}
which does not seem to be in a gross contradiction with
\cite{ChCh95}. In other words, 3 phases of the Polyakov loop ``mix
together'' below $T_c$.

In a picture of the chiral phase transition in the quenched QCD
suggested here the restoration of the chiral symmetry at finite $T$ is
mainly due to the effect of the b.c. in the Euclidean time suppressing
small eigenvalues of the Dirac operator.  The rearrangement of the QCD
equilibrium state plays a secondary role. This goes in line with the
fact that, as far as we know, the instanton density behaves rather
smoothly across the phase
transition~\cite{Sh95,ChSc94,PiWi84}.\footnote{We must emphasize here
that the present discussion mainly concerns a model of QCD where the
feedback of quarks onto the dynamics of gluons is absent. The idea
that confinement is not a prerequisite of the chiral symmetry breaking
has been also suggested by a study~\cite{KoSt83} of the $SU(2)$ theory
with adjoint quarks.} The effect of the b.c. in the $\arg P=0$ phase
is stronger than in the $\arg P=2\pi/3$ phase and in the low
temperature $P=0$ phase. To put this into another perspective one can
note that antiperiodic boundary conditions for the quark fields are
related to the Pauli exclusion principle.\footnote{For a discussion of
the relation between the Pauli principle and the decoupling of
fermions from a dimensionally reduced theory at high $T$ see
\cite{St95}.} The fact that Pauli blocking leads to chiral symmetry
restoration is rather well known in various models of particle and
nuclear physics.
 
Another interesting consequence of (\ref{tchi}) is that in the $SU(2)$
gauge theory: $T_\chi(\pi)=\infty$. This means that the
$\langle\bar\psi\psi\rangle$ stays nonzero for all $T$ in the $\arg
P=\pi$ phase!

This result can be argued for also beyond the approximation of the
random matrix model. Recall that the theory at very high $T\gg T_c$
becomes effectively 3-dimensional. Moreover, fermions which would
``decouple'' in the $\arg P=0$ phase do not in the $\arg P=\pi$ phase
because the b.c. are periodic. Thus the value of
$\langle\bar\psi\psi\rangle$ is determined by the 3-dimensional
$SU(2)$ gauge theory at {\em zero} temperature. The value of the
$\langle\bar\psi\psi\rangle$ is nonzero in such a
theory~\cite{DaKo91}. Moreover, it can be estimated to be of the order
of $g_3^4$ purely on dimensional grounds. This means that the chiral
condensate in the $SU(2)$ theory in the $\arg P=\pi$ phase at $T\gg
T_c$ is of the order:
\begin{equation}\label{pbpsu2}
\langle\bar\psi\psi\rangle\sim g^4T^3.
\end{equation}

Recently there have been two attempts \cite{MeOg95,ChHu95} to describe
the data \cite{ChCh95} using a Nambu-Jona-Lasinio model. With some
tuning of parameters it is possible to achieve a reasonable fit to the
data in the model \cite{ChHu95}. The present paper emphasizes the role
of the behavior of small eigenvalues as a function of $\arg P$ and in
this respect differs from a somewhat phenomenological approach of
\cite{MeOg95,ChHu95}. This allows us to obtain new results, for
example,~(\ref{tchi}) or~(\ref{pbpsu2}).

It would be interesting to test these results in a Monte Carlo
simulation. For example, one can consider b.c. conditions on fermions
with an arbitrary twist. This would add to our knowledge of the mechanism
of the chiral symmetry restoration in QCD.

{\bf Acknowledgments: }\ The author would like to thank A. Kocic,
J. Kogut, C. Korthals-Altes and R. Pisarski (who brought the result
\cite{ChCh95} to the author's attention) for useful discussions and
comments and the theory group of the Brookhaven National Laboratory
for hospitality. The work was supported by the NSF grant
PHY~92--00148.

\appendix

\section{The density of eigenvalues}
The eigenvalue density $\rho(\lambda)$ for the Dirac operator can be
found in a closed form in our approximation. 
Let us consider only the lowest band of eigenvalues with
$\omega=\min_n|\omega_n|$. Contributions of the other bands are additive. To
find $\rho(\lambda)$ we note that one can calculate
$\langle\bar\psi\psi\rangle(m)$ for arbitrary $m$.
On the other hand,
$\rho(\lambda)$ is the discontinuity of
$\langle\bar\psi\psi\rangle(m)$ along the imaginary axis, in other
words:
\begin{equation}
\mbox{ Re }\langle\bar\psi\psi\rangle(i\lambda+\epsilon)
=\pi\rho(\lambda).
\end{equation}

The $\langle\bar\psi\psi\rangle(m)$ is given by (\ref{pbptr}) if we do
not take $m\to0$. The necessary eigenvalue $y$ can be found
from~(\ref{saddle}) by solving a cubic equation. Finally, we obtain:
\begin{equation}
{\pi\rho(\lambda)\over C}={\sqrt3\over2}\left[
({\sqrt \Delta} + q)^{1/3} + 
({\sqrt \Delta} - q)^{1/3} \right],
\end{equation}
where one should take the branch $(-1)^{1/3}=-1$, and
\begin{equation}
q={\tilde\lambda\over6}
\left(1+2\tilde\omega^2-{2\over9}\tilde\lambda^2\right),
\qquad
\Delta={1\over27}\left[
(1-\tilde\omega^2)^3 - {\tilde\lambda^2\over4}(1-20\tilde\omega^2-
8\tilde\omega^4 + 4\tilde\lambda^2\tilde\omega^2)
\right],
\end{equation}
where $\tilde\lambda=\lambda C$ and $\tilde\omega=\omega C$.

\begin{figure}[hbt]
                \centerline{    \epsfysize=2.4in
                                \epsfbox{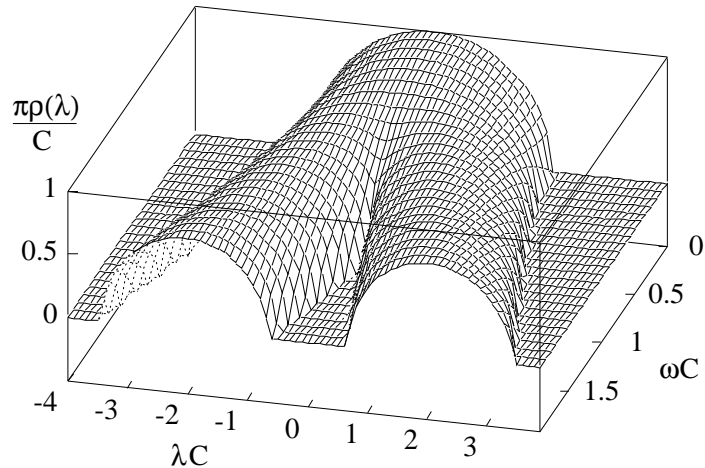}
                           }
\bigskip
\caption[]{The eigenvalue density of the Dirac operator for 
a single band $\omega$ as a function of $\omega$.}
\label{fig:rho}
\end{figure}

The shape of this distribution is shown in Fig.~\ref{fig:rho}. At
$\omega=0$ the $\rho(\lambda)$ has a semicircular shape.  As $\omega$
increases the distribution develops a dip at $\lambda=0$ and when
$\omega>(1/C)$ it splits into two bands leaving $\rho(\lambda=0)=0$.

\begin{thebibliography}{99}
\bibitem{BaCa80} 
		T. Banks and A. Casher, Nucl. Phys. B169 (1980) 103.
\bibitem{Sh95} 
		E.V. Shuryak, 
		hep-ph/9503427, 
		to be published in 
		{\it Quark-Gluon Plasma}, ed. R. Hwa. 
\bibitem{ChCh95}  
		S. Chandrasekharan and N. Christ, Contribution 
		to International Symposium on
		Lattice Field Theory, Melbourne, Australia, 11-15 Jul 1995,
		hep-lat/9509095.
\bibitem{KoSt83}
		J.B. Kogut et al, 
		Phys. Rev. Lett. 50 (1983) 393;
		Nucl. Phys. B225 (1983) 326.
\bibitem{Po77} 
		A.M. Polyakov, Phys. Lett. 72B (1977) 477.
\bibitem{hist}
		Y. Iwasaki et al, Phys. Rev. Lett. 67 (1991) 141;
		M.A. Stephanov and M.M. Tsypin,
		Nucl. Phys. B 366 (1991) 420.
\bibitem{GrPi81}
		D.J. Gross, R.D. Pisarski and L.G. Yaffe,
		Rev. Mod. Phys. 53 (1981) 43.
\bibitem{ShVe93}
		E.V. Shuryak and J.J.M. Verbaarschot,
		Nucl. Phys. A560 (1993) 306. 
\bibitem{JaVe95}
		A.D. Jackson and J.J.M. Verbaarschot,
		hep-ph/9509324.
\bibitem{ChSc94}
		M.C. Chu and S. Schramm, 
		Phys. Rev. D51 (1995) 4580.
\bibitem{PiWi84} 
                R. Pisarski and F. Wilczek, 
		Phys. Rev. D29 (1984) 338.
\bibitem{DaKo91}
		E. Dagotto, A. Kocic and J.B. Kogut, 
		Nucl. Phys. B 362 (1991) 498
\bibitem{St95}
		M.A. Stephanov,
		Phys. Rev. D52 (1995) 3746.
\bibitem{MeOg95}
		P.N. Meisinger and  M.C. Ogilvie, hep-lat/9512011.
\bibitem{ChHu95}
		S. Chandrasekharan and S. Huang, hep-ph/9512323.
\end{thebibliography}
\end{document}